\documentclass[lettersize,journal]{IEEEtran}

\usepackage{amsmath,amsfonts}
\usepackage[font=small,justification=centering]{caption}
\usepackage{algorithm}
\usepackage{algorithmicx}
\usepackage{algpseudocode}

\usepackage{array}
\usepackage[caption=false,font=normalsize,labelfont=sf,textfont=sf]{subfig}
\usepackage{textcomp}
\usepackage{stfloats}
\usepackage{url}

\usepackage{verbatim}
\usepackage{graphicx}
\usepackage{float}     
\usepackage{xcolor}
\usepackage[table,xcdraw]{xcolor}

\usepackage{longtable}
\usepackage{booktabs}
\usepackage{multirow}
\usepackage{lscape}
\usepackage{graphicx}
\usepackage{cite}
\usepackage{caption} 

\begin{document}

\title{Fuzzychain-edge: A novel Fuzzy logic-based adaptive Access control model for Blockchain in Edge Computing}

\author{Khushbakht Farooq, Muhammad Ibrahim,Irsa Manzoor, Mukhtaj Khan~\IEEEmembership{Senior Member,~IEEE}, and Wei Song ~\IEEEmembership{Fellow,~IEEE}
\thanks{This paper was produced by the IEEE Publication Technology Group. They are in Piscataway, NJ.}
\thanks{Manuscript received April 19, 2021; revised August 16, 2021.}}

\markboth{Journal of \LaTeX\ Class Files,~Vol.~14, No.~8, August~2021}%
{Shell \MakeLowercase{\textit{et al.}}: A Sample Article Using IEEEtran.cls for IEEE Journals}

\IEEEpubid{0000--0000/00\$00.00~\copyright~2021 IEEE}

\maketitle

\begin{abstract}
The rapid integration of internet of things(IoT) with edge computing has 
revolutionized various domains, particularly healthcare, by enabling real-time data sharing, remote monitoring, and decision-making. However, it introduces critical challenges, including data privacy breaches, security vulnerabilities, and limited traceability, especially in environments dealing with sensitive information. Traditional access control mechanisms and centralized security systems do not address these issues, leaving IoT environments exposed to unauthorized access and data misuse. This research proposes Fuzzychain-edge, a novel Fuzzy logic-based adaptive Access control model for Blockchain in Edge Computing framework designed to overcome these limitations by incorporating Zero-Knowledge Proofs (ZKPs), fuzzy logic, and smart contracts. ZKPs secure sensitive data during access control processes by enabling verification without revealing confidential details, thereby ensuring user privacy. Fuzzy logic facilitates adaptive, context-aware decision-making for access 
control by dynamically evaluating parameters such as data sensitivity, trust levels, and user roles. Blockchain technology, with its decentralized and immutable architecture, ensures transparency, traceability, and accountability using smart contracts that 
automate access control processes. The proposed framework addresses key challenges by enhancing security, reducing the likelihood of unauthorized access, and providing a transparent audit trail of data transactions. Expected outcomes include improved 
data privacy, accuracy in access control, and increased user trust in IoT systems. This 
research contributes significantly to advancing privacy-preserving, secure, and 
traceable solutions in IoT environments, laying the groundwork for future innovations in decentralized technologies and their applications in critical domains such as healthcare and beyond.
\end{abstract}
\section{Introduction}

Internet of Things (IoT) is a revolutionary technology that provides interaction between ordinary devices with the ability to gather data, exchange it, and analyze it in an efficient manner. IoT devices on the one hand produce enormous amounts of real-time data used to make timely decisions and automation whether in the healthcare monitoring industry or the industrial sector in general \cite{1,2}. Although cloud computing offers the required computing power to process this data, its centralized application presents a problem of latency, which is not appropriate in applications that are sensitive to delay as cloud computing is not suitable at all \cite{3}. To overcome these limitations, edge and fog computing paradigms have emerged, bringing computational resources closer to data sources and reducing response times for critical IoT services \cite{4,5}.

Security and privacy are also hot topics in tandem with these developments. IoT devices can be vulnerable to unauthorized access and cyberattacks because sensitive data can be provided, and the information can be personal health data, etc.\cite{6}.  A solution to this problem is blockchain technology that provides decentralized, tamper resistant, and transparent management of data. The cryptographically linked block structure ensures immutability and traceability without the use of trusted third parties and provides reliable records of transactions and accessibility to transactions and accessibility to records of transactions and accessibility to records of transactions and accessibility to records of transactions and accessibility to records of transactions and accessibility to records of access\cite{7,8}. 

The use of blockchain in IoT can ensure safe and auditable control of access. Decentralized schemes such as the Attribute-Based Access Control (ABAC), Role-Based Access Control (RBAC), and Capability-Based Access Control (CBAC) enable the fine-tuning of user permissions, which is in line with the transparency and immutability characteristics of blockchain in nature \cite{9,10}. To further strengthen privacy, techniques such as Zero-Knowledge Proofs (ZKP) enable users to prove their authorization or validate data without revealing sensitive information \cite{11,12}. This is especially useful in the medical and the financial sector where confidentiality is critical.
Fuzzy logic complements these mechanisms by providing adaptive, context-aware decision-making in access control, allowing systems to handle uncertainties and varying conditions dynamically \cite{13}.  This study will use a combination of blockchain, ZKP, and fuzzy logic to establish a sound framework that will guarantee secure, privacy-sensitive, and trustworthy IoT activities. The suggested framework, in addition to improving the access control, provides traceability, accountability, and effective management of the IoT services.

This research addresses the challenge of ensuring secure, privacy-preserving, and efficient access control in IoT systems that use blockchain, and the overall aim of this study is to secure and protect user data, promote transparency in the system, and allow reliable working of this IoT. To achieve this, we propose a new framework to combine blockchain technology with zero-knowledge proofs (ZKP), access control based on fuzzy logic, and traceability supported by smart contracts.  The framework provides secure access, upholds privacy, and accountability by dynamically assessing user attributes, access permissions, and system conditions to ensure the safety of access to the decentralized IoT networks. The results of the experiment prove that the offered framework can be improved in terms of security, privacy, and efficient management of IoT services in comparison with the current strategies. Specifically, the main contributions of this research are as follows:
\begin{itemize}
   
    \item Access control model based on blockchain with the integration of Zero-Knowledge Proofs (ZKP) to maintain user privacy and secure verification of user access rights..
    
    \item A fuzzy logic based decision system, which allows adjusting and autonomous assigning and revoking roles, enhancing access control in dynamic IoT systems.
    
    \item A traceability module, powered by smart contract and able to record and audit all access requests, which increases accountability, transparency, and trust in the activities of IoT.
    
    \item The comparison of the proposed approach with the current practices can be used to evaluate the performance of the approach with regard to privacy, security, and efficiency of decentralized IoT applications.
\end{itemize}

The rest of the paper is organized as follows: Section II reviews related work; Section III presents the proposed methodology and framework; Section IV describes experimental results and performance evaluation; and Section V concludes the study and discusses future research directions.\\ 

\section{Related Work}
\label{RW}

In this section, we present a comprehensive review of the literature focusing on privacy preservation, security in terms of Access Control, and traceability within blockchain-based healthcare IoT systems, which form the basis of this research. Furthermore, it discusses more recent analyses that encompass current methods for data protection, access control, and accountability in decentralized IoT structures.

The Internet of Things (IoT) devices have rapidly proliferated in recent years across many different sectors, such as smart cities and healthcare. However, addressing privacy, security, access control, and traceability in the context of 
integrating these devices into everyday life is a very difficult problem. With the growing interconnectivity of IoT systems, there is a growing potential for data breaches, unauthorized access, and privacy violations. These characteristics of 
IoT environments are not addressed by traditional security measures, which tend to be resource-consuming, use a homogeneous architecture, and require real-time processing \cite{14}. 

With the advent of blockchain technology, these challenges have a promising solution. The security and privacy of IoT systems can be improved by leveraging blockchain’s decentralized and immutable nature, while maintaining a transparent audit trail of data transactions. The aim of this literature 
review is to look at the interface between blockchain technology and IoT and more specifically, the mechanisms that improve privacy, security, traceability, 
and access control \cite{15}. 

This review covers a range of dimensions of blockchain-integrated IoT security, including data integrity and traceability through blockchain, privacy preservation 
through Zero Knowledge proofs (ZKPs), fuzzy logic in access control models and smart contracts as a tool of automated access management. This review synthesizes the current state of research in these areas and attempts to identify existing frameworks, discuss their shortcomings, and identify gaps in the literature that should be further explored \cite{16}. To address security and privacy concerns in decentralized IoMT systems, the authors proposed the Fortified-Chain framework, which is based on blockchain technology. 

To address challenges such as high latency, high cost of storing data, and having a single point of failure in cloud centric IoMT systems, the paper suggests a hybrid system that integrates blockchain with a Distributed Data Storage System (DDSS). Selective Ring-based Access Control SRAC integrated at the architecture level, along with device authentication and anonymous patient record algorithms, assures secure anonymous data accesses. Some smart services basically include electronic health records management, where the services are automated through the smart contract hence increasing security and privacy. Based on private blockchains and tailored for low-latency and inexpensive 
information exchange, the framework is assessed through experiments yielding submillisecond response times. In theoretical analysis with logical reasoning and simulation, it has been proved that the present approach is scalable and efficient. \cite{17}.

Zero-Knowledge Proofs (ZKPs) are cryptographic protocols through which one party, the so-called prover, can convince another party, the verifier, that a certain statement is correct without disclosing any information about the statement. This idea revolutionizes privacy in IoT access control by permitting the devices to identify themselves and exchange information that is relevant without necessarily having to expose other information that is equally relevant but considered sensitive. For instance, ZKPs can be applied in contexts where an IoT device must prove that it follows the use of security standards without exposing the inner settings, or users’ data. This capability is especially valuable in areas where data security is a priority like the health sector and the financial sector because it lowers the chances of data leakage while at the same time confirming the permissions of devices \cite{18,19}. Proofs of current applications of ZKPs in blockchain and IoT are given to show that this technology is valuable in practice. Some of these include the application of ZKPs in cryptocurrency where they add security on user’s anonymity by hiding transaction details but can prove their authenticity. Furthermore, using ZKPs, projects like Zcash make use of the functionality to facilitate the operation of shielded transactions also \cite{20}.

This work proposes a blockchain-based zero-knowledge model for secure data sharing and access control. ZKPs are employed to verify user permissions while concealing identity and data attributes. Blockchain ensures decentralization and tamper resistance. The model improves privacy and trust in data exchange. Nevertheless, computational complexity and scalability limitations are not fully addressed\cite{21}. This paper introduces a privacy-preserving account-based blockchain using zk-SNARKs to overcome transaction linkability issues in Ethereum-like systems. The scheme hides account balances and transaction relationships. Formal security proofs and implementation results validate strong privacy protection. The work achieves a balance between privacy and efficiency. However, transaction generation time remains relatively high\cite{22}. This paper present a blockchain-based IoT access control model using zero-knowledge tokens. The model protects user identity and access attributes through encrypted tokens stored on the blockchain. Smart contracts manage authorization without relying on centralized servers. Experimental results show improved scalability compared to traditional ABAC systems. The approach is suitable for decentralized IoT environments\cite{23}.

This paper presents a privacy-preserving healthcare IoT system based on blockchain and zk-SNARKs. Sensitive medical data are protected through anonymous verification mechanisms. The system resists common attacks such as impersonation and man-in-the-middle attacks. Performance evaluation shows moderate computational overhead. The solution is effective for secure smart healthcare applications\cite{24}.This work proposes a blockchain-based medical insurance claim framework using zero-knowledge proof technology. Secure verification ensures privacy, authenticity, and anonymity during insurance and claim processes. Non-interactive ZKPs and cryptographic techniques protect sensitive data. Security analysis confirms robustness against data leakage. The framework demonstrates practical feasibility\cite{25}. A zero-knowledge-proof-based privacy-preserving mutual authentication scheme for IoT networks is introduced. A permissioned blockchain is employed to achieve low latency and data integrity. Device identities are protected during authentication using ZKPs. Experimental results show significant reduction in authentication time. The scheme balances security, scalability, and privacy\cite{26}.

This study aims to fill the gaps that characterize current disease reporting systems, including central control, data loss, and questionable traceability, by presenting a blockchain system for infectious disease traceability. Based on the blockchain characteristics of decentralization, non-tamperability, and transparency of the whole process, the system realizes the safe and efficient collection and tracking of disease information. Components are information gathering, a chain-like structure, and querying. The proposed system creates a time series of blockchain of disease data, making the data authentic and transparent. With respect to infectious diseases, it enables tracking of disease transmission paths while it enhances real-time monitoring and querying of disease information to increase reliability, traceability, and usefulness of responses to help control disease spreading\cite{27}. 
 
Simulations prove the increase in the system’s ability to support enhanced traceability of work and the increased defense of primary accomplishments. Moreover, at the same time the system has an automatic incentive function to encourage creation and protection activities. Comparative analysis is done to show benefits involved in using this blockchain approach over the traditional procedures. The paper also outlines numerous associated general applications of blockchain across various industries apart from presenting a concrete solution to the problem of IPR protection\cite{28}.
This section provides a comprehensive review of literature focusing on privacy preservation, security in terms of Access Control and traceability within blockchain based healthcare IoT systems, which form the basis of this research. Furthermore, it discusses more recent analyses that encompass current methods for data protection, access control, and accountability in decentralized IoT structures.

\begin{table*}[htbp]
\caption{Comparative Analysis of Result Evaluation Parameters}
\label{table:result_comparison}
\centering
\resizebox{\textwidth}{!}{
\begin{tabular}{|l|c|c|c|c|c|c|c|}
\hline
\textbf{Work} & \textbf{Latency} & \textbf{Throughput} & \textbf{Send Rate} & \textbf{Block Size} & \textbf{Privacy} & \textbf{Access Control} & \textbf{Scalability} \\
\hline
Said (2022) [14] & \checkmark & \checkmark & \texttimes & \texttimes & \checkmark & \checkmark & \texttimes \\
Haritha \& Anitha (2023) [15] & \checkmark & \checkmark & \texttimes & \texttimes & \checkmark & \checkmark & \texttimes \\
Kumar et al. (2022) [16] & \checkmark & \checkmark & \texttimes & \texttimes & \checkmark & \checkmark & \checkmark \\
Chen et al. (2021) [17] & \checkmark & \checkmark & \texttimes & \texttimes & \checkmark & \checkmark & \texttimes \\
Egala et al. (2021) [18] & \checkmark & \checkmark & \texttimes & \texttimes & \checkmark & \checkmark & \checkmark \\
Yang \& Li (2020) [19] & \checkmark & \texttimes & \texttimes & \texttimes & \checkmark & \texttimes & \texttimes \\
Saroop (2024) [20] & \checkmark & \checkmark & \texttimes & \texttimes & \checkmark & \checkmark & \texttimes \\
Al-Aswad et al. (2019) [21] & \checkmark & \texttimes & \texttimes & \texttimes & \checkmark & \texttimes & \texttimes \\
Guan et al. (2022) [22] & \checkmark & \checkmark & \texttimes & \texttimes & \checkmark & \checkmark & \texttimes \\
Song et al. (2021) [23] & \checkmark & \checkmark & \texttimes & \texttimes & \checkmark & \checkmark & \texttimes \\
Luong \& Park (2022) [24] & \checkmark & \checkmark & \texttimes & \texttimes & \checkmark & \checkmark & \texttimes \\
Zheng et al. (2022) [25] & \checkmark & \checkmark & \texttimes & \texttimes & \checkmark & \checkmark & \texttimes \\

\textbf{Proposed Model} & \textbf{\checkmark} & \textbf{\checkmark} & \textbf{\checkmark} & \textbf{\checkmark} & \textbf{\checkmark} & \textbf{\checkmark} & \textbf{\checkmark} \\
\hline
\end{tabular}
}
\end{table*}

\color{black}
\section {System Model}
\label{prop}
The system architecture involve fuzzy logic, zero-knowledge proofs (zk- SNARKs) 
and a blockchain system for enhanced access control over healthcare information 
systems. Users first submit their credentials consisting of roles, subject attributes, and 
object attributes, where the information is proactively protected through zk-SNARK. 
These credentials are encrypted and stored on a blockchain database so that all 
verifiable information is safe and cannot be altered. In an access request scenario, 
users’ zk-identity, along with their data, are treated and forwarded to the subsequent 
validation step. Zero-knowledge proofs used in this scenario ensures that the identity 
of the user is kept secret while at the same time verifying the data.

The main component of the access control system is the Fuzzy Logic module, which is 
aimed at real-time processing of the access requests. The capability token produced by 
the module is (step 3.1), and fuzzy-based rules (step 3.2) are used to make access 
decisions with respect to the parameters like data sensitivity, user role, and trust level. 
This approach increases flexibility as well as putting more emphasis on the fact that 
access control decisions should change depending on the current or required request. 
The access control module then produces an output that is forwarded for decision-making while delivering an overall analysis prior to allowing access or denying it. 
Blockchain is used to ensure that all the decisions made are traceable, hence providing 
an audit trail of the access activities.

\begin{figure*}[!t]
    \centering
   \includegraphics[width=\textwidth]{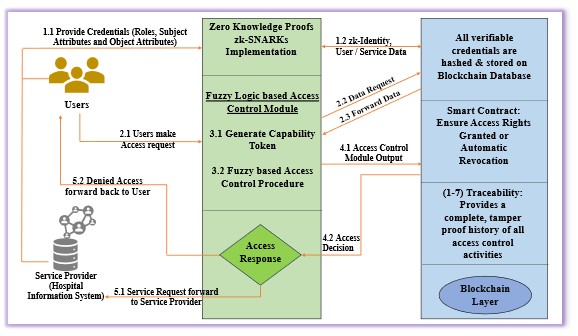}

    \caption{System architecture of proposed Fuzzychain-edge framework}
    \label{fig:System model}
\end{figure*}

Last, in access responses, the messages are conveyed to the service provider and the 
user. If access is granted, all the rights are managed through the smart contract on the 
blockchain, guaranteeing the requested services’ security. In case of denial, the 
response is taken back to the user with the reason for denial, making the system 
transparent. The blockchain layer also preserves the uniqueness and traceability 
record of all the access control activities throughout steps 1 to 7, making it possible to 
enhance the reliable and secure revocation and granting of rights in the development 
of a healthcare information system. This methodology as shown in Figure~\ref{fig:System model}
provides a strong solution to privacy, security, and efficiency issues in sensitive areas 
such as healthcare.

\subsection{Definitions and Initialization}

Let: 
CU=(IDU, AU): User credentials, where AU is a vector of user attributes (SA,OA,TL,CH,…). 

CS=(IDS, AS): Service provider credentials, where AS is a vector of service 
attributes (SA,OA,DS,RA,…).

kp , vk: Proving and verification keys for zk-SNARKs. 

f: The proving function that generates zk-SNARK proof. 

V : The verification function that checks the zk-SNARK proof. 

H(x): A cryptographic hash function, e.g., SHA-256.

B: The blockchain, storing hashed records.

\subsection{zk-SNARK Proof Generation}

For a proving function \(f\), generate zk-SNARK proofs for the service provider and user credentials:
\begin{equation}
P_S = f(CS, k_p), \quad P_U = f(CU, k_p)
\end{equation}
Where \(P_S\) and \(P_U\) are succinct proofs encapsulating the validity of \(CS\) and \(CU\) under zk-SNARK.  

\subsection{zk-SNARK Proof Verification}

For a verification function \(V\), verify the generated proofs using the verification key \(v_k\):
\begin{equation}
V_S = V(P_S, v_k), \quad V_U = V(P_U, v_k)
\end{equation}

Conditions:
\[
V_S = 
\begin{cases}
\text{True}, & \text{if } P_S \text{ is valid} \\
\text{False}, & \text{otherwise}
\end{cases}
,\quad
V_U =
\begin{cases}
\text{True}, & \text{if } P_U \text{ is valid} \\
\text{False}, & \text{otherwise}
\end{cases}
\]

\textbf{Advanced Representation Using Formal Logic}

\begin{equation}
\forall i \in \{S, U\}, \quad V(P_i, v_k) \implies \text{VerificationSuccess}
\end{equation}

If any proof fails:
\begin{equation}
\exists i \in \{S, U\}, \quad \neg V(P_i, v_k) \implies \text{VerificationFailure}
\end{equation}

If \(V_S = \text{False}\), terminate with:
\[
\text{"ServiceProviderVerificationFailed"}
\]

If \(V_U = \text{False}\), terminate with:
\[
\text{"UserVerificationFailed"}
\]

\subsection{Privacy-Preserving Data Processing}

Transform the data (\(CS\) or \(CU\)) into a privacy-preserved format using a secure function \(P\):
\begin{equation}
\text{PrivacyPreservedData}_S = P(CS), \quad \text{PrivacyPreservedData}_U = P(CU)
\end{equation}

\subsection{Cryptographic Hashing and Blockchain Storage}

Hash the privacy-preserved data for immutable storage in the blockchain:
\begin{equation}
\text{Blockchain}_S = H(\text{PrivacyPreservedData}_S)
\end{equation}
\begin{equation}
\text{Blockchain}_U = H(\text{PrivacyPreservedData}_U)
\end{equation}

Store the hashes in the blockchain \(B\):
\begin{equation}
B \leftarrow B \cup \{\text{Blockchain}_S, \text{Blockchain}_U\}
\end{equation}

\subsection{zk-Identity Generation}

Derive zk-Identities for the service provider and user based on their hashed, privacy-preserved data:
\begin{equation}
kID_S = H(\text{Blockchain}_S \oplus TS)
\end{equation}
\begin{equation}
kID_U = H(\text{Blockchain}_U \oplus TU)
\end{equation}
Where \(TS\) and \(TU\) are unique salt values for the service provider and user, ensuring unlinkability.

\subsection{Output and Final Verification}

Define the Final Verification condition as:
\begin{equation}
\text{FinalVerification} = V_S \wedge V_U
\end{equation}

Return outputs based on Final Verification:
\[
\text{Output} =
\begin{cases}
\text{zk-IdentityDataStoredSuccessfully}, & \text{if 
FinalVerification = True} \\
\text{ServiceProviderVerificationFailed}, & \text{if } V_S = \text{False} \\
\text{UserVerificationFailed}, & \text{if } V_U = \text{False}
\end{cases}
\]
\begin{table}[!t]
\caption{Notation used in the blockchain-based access control system}
\label{table:notation_blockchain}
\centering
\scriptsize
\begin{tabular}{ll}
\toprule
\textbf{Parameter Description} & \textbf{Symbol} \\
\midrule
User credentials & $CU = (ID_U, AU)$ \\
User attribute vector & $AU = (SA, OA, TL, CH, \dots)$ \\
Service provider credentials & $CS = (ID_S, AS)$ \\
Service attribute vector & $AS = (SA, OA, DS, RA, \dots)$ \\
Proving key for zk-SNARK & $kp$ \\
Verification key for zk-SNARK & $vk$ \\
Proving function & $f$ \\
Verification function & $V$ \\
Privacy-preserving function & $P$ \\
Cryptographic hash function & $H(x)$ \\
Blockchain & $B$ \\
zk-SNARK proof of user & $P_U$ \\
zk-SNARK proof of service provider & $P_S$ \\
Verification result of user & $V_U$ \\
Verification result of service provider & $V_S$ \\
zk-Identity of user & $kID_U$ \\
zk-Identity of service provider & $kID_S$ \\
Unique salt for user & $T_U$ \\
Unique salt for service provider & $T_S$ \\
Capability token from fuzzy logic & $C(\tau_i, \nu_j)$ \\
Access compliance & $AC_i$ \\
Blockchain logging status & $BL_i$ \\
Final access decision & $AD_i$ \\
Average response time & $\varphi$ \\
Total number of access requests & $n$ \\
\bottomrule
\end{tabular}
\end{table}

\section{PROPOSED ALGORITHMS}
\label{exp-eval}
\subsection{Access Control Hub}

In Algorithm 1, the inputs at the beginning of the algorithm are the Subject Attributes 
(SA), Object Attributes (OA), Roles, Service Hash Table (ServiceHash) and the User Hash Table (UserHash). These components are very important in providing the distinctive identification of clients, services, and the other attributes used for access control purposes. The main result of the application of this algorithm is the zk-Identity of users and services, which are kept on the blockchain for record and identification. 
First, the initialization of the Service and User Registration is called (line 2). In this 
function, the algorithm first initializes the two hash tables of ServiceHash and UserHash with empty forms (line 3). These tables will be later filled with hashed and encrypted data which will serve as storage for attributes. 
Next, the algorithm collects a set of service-related attributes (CS) such as the Service 
ID, Subject Attributes (SA), Object Attributes (OA), Trust Level (TL), Priority Class 
(PC), Data Sensitivity (DS), Resource Availability (RA), Access Priority (AP), 
Treatment Urgency (TU), and Compliance History (CH) (line 4). These attributes are 
passed through a zk-SNARK proof generation and verification process to ensure their 
authenticity and validity (line 5). Once verified, a Zero-Knowledge Identity (zkIDS) is 
generated for the service attributes (line 6). This identity is then hashed along with 
service data HBlockchainS←HashzkIDS,DataS (line 7), ensuring privacy and immutability. The hashed service data is stored on the blockchain (line 8). 
Similarly, the algorithm collects a set of user-related attributes (CU) such as the User ID, SA, OA, TL, PC, DS, RA, AP, TU, and CH (line 9). These user attributes undergo the same zk-SNARK proof generation and verification process to validate their 
authenticity (line 10). A Zero-Knowledge Identity (zkIDU) is then generated for the 
user attributes (line 11). The algorithm hashes the user identity along with the user’s 
data HBlockchainU←HashzkIDU,DataU (line 12), and this hashed information is stored on the blockchain for secure access and tamper-proof verification (line 13). 
Finally, the algorithm evaluates the Degree of Access (DoA) to determine whether access should be granted or denied. If the DoA is greater than zero (line 14), the 
algorithm returns “Access Allowed” (line 15), enabling the user to access the requested service. Otherwise, if the DoA is zero or less, the algorithm returns “Access 
Denied” (lines 17–18), denying the access request and ensuring that unauthorized or insufficiently verified entities cannot proceed.

\begin{algorithm}
\caption{Access Control Hub}
\begin{algorithmic}[1]

\State \textbf{Input:} Subject Attributes ($SA$), Object Attributes ($OA$), Roles,
Service Hash Table ($ServiceHash$), User Hash Table ($UserHash$),
zk-Identity user/service data stored on blockchain

\Function{ServiceAndUserRegistration}{}
    \State Initialize $ServiceHash$ and $UserHash$ as empty hash tables
    \State Collect $C_S = \{ServiceID, SA, OA, TL, PC, DS, RA, AP, TU, CH\}$
    \State Generate and verify zk-SNARK proof for $C_S$
    \State $zkID_S \leftarrow GenerateZKIdentity(C_S)$
    \State $H_{\text{Blockchain}_S} \leftarrow \text{Hash}(zkID_S, Data_S)$
    \State Store $H_{\text{Blockchain}_S}$ in blockchain database

    \State Collect $C_U = \{UserID, SA, OA, TL, PC, DS, RA, AP, TU, CH\}$
    \State Generate and verify zk-SNARK proof for $C_U$
    \State $zkID_U \leftarrow GenerateZKIdentity(C_U)$
    \State $H_{\text{Blockchain}_U} \leftarrow \text{Hash}(zkID_U, Data_U)$
    \State Store $H_{\text{Blockchain}_U}$ in blockchain database

    \If{$DoA > 0$}
        \State \Return Access Allowed
    \Else
        \State \Return Access Denied
    \EndIf
\EndFunction

\end{algorithmic}
\end{algorithm}

\subsection{Zero Knowledge Proof (zk-SNARK) Process}
  
The algorithm 2 given below process first defines the input and the output. The inputs 
include user credentials (CU), service provider credentials (CS), as well as 
cryptographic keys required for zk-SNARK operations: Two of the keys that are used 
are a ProvingKey and a VerificationKey. The aim is to derive peer-reputed zk-identity 
data for the services and entities to be archived in the blockchain network. This makes the system private and unalterable while at the same time putting in place secure and 
accountable access controls. 
The algorithm is designed to securely validate and store the identities of both service 
providers and users on the blockchain using zk-SNARKs. The inputs include User 
Credentials (CU), Service Provider Credentials (CS), a Proving Key, and a Verification 
Key. These keys facilitate the generation and verification of zero-knowledge proofs, 
ensuring data authenticity and privacy. The primary goal is to ensure that the zk
identity data of both entities is securely stored on the blockchain. The process begins 
with the zk-SNARK Proof Generation and Verification Function (line 2). First, the 
algorithm generates a zero-knowledge proof zkProofS←ProveCS,ProvingKey; for the 
service provider using the provided service credentials and the Proving Key (line 3). 
This proof validates the service provider's identity without exposing sensitive details. 
The proof is then verified using the Verification Key to ensure its authenticity 
verificationS←VerifyzkProofS,VerificationKey (line 4).  
If the service provider's proof verification is successful verificationS=True (line 5), the 
algorithm 
processes 
the 
service 
provider's 
data 
privacy 
privacyPreservedDataS←ProcessDataCS to ensure sensitive information is protected 
(line 6). This processed data, along with the service provider’s zk-identity proof, is 
hashed HBlockchainS ←HashprivacyPreservedDataS (line 7). The resulting hash is 
then stored securely in the blockchain database, providing an immutable record of the 
service provider’s verified identity and attributes (line 8). If the proof verification 
fails, the algorithm returns “Service Provider Verification Failed” (lines 10–11), 
indicating that the service provider's credentials could not be validated. 
 
\begin{algorithm}
\caption{Zero Knowledge Proof (zk-SNARK) Process with zk-identity Service Provider, Users, and Data Storage on Blockchain}
User Credentials ($C_U$), Service Provider Credentials ($C_S$), ProvingKey,
VerificationKey Verified Service Provider and Users zk-identity data stored on Blockchain
\begin{algorithmic}[1]

\Function{zkSNARKProofGenerationAndVerification}{}

    \State $zkProof_S \leftarrow \text{Prove}(C_S, ProvingKey)$
    \State $verification_S \leftarrow \text{Verify}(zkProof_S, VerificationKey)$

    \If{$verification_S = \text{True}$}
        \State $privacyPreservedData_S \leftarrow \text{ProcessData}(C_S)$
        \State $H_{\text{Blockchain}_S} \leftarrow \text{Hash}(privacyPreservedData_S)$
        \State Store $H_{\text{Blockchain}_S}$ in blockchain database
    \Else
        \State \Return Service Provider Verification Failed
    \EndIf

    \State $zkProof_U \leftarrow \text{Prove}(C_U, ProvingKey)$
    \State $verification_U \leftarrow \text{Verify}(zkProof_U, VerificationKey)$

    \If{$verification_U = \text{True}$}
        \State $privacyPreservedData_U \leftarrow \text{ProcessData}(C_U)$
        \State $H_{\text{Blockchain}_U} \leftarrow \text{Hash}(privacyPreservedData_U)$
        \State Store $H_{\text{Blockchain}_U}$ in blockchain database
        \State \Return zk-Identity Data Stored Successfully
    \Else
        \State \Return User Verification Failed
    \EndIf

\EndFunction

\end{algorithmic}
\end{algorithm}
The algorithm proceeds similarly for the user. It generates a zero-knowledge proof 
(zkProofU←ProveCU,ProvingKey) for the user using their credentials (CU) and the 
Proving Key (line 13). The proof is verified with the Verification Key 
(verificationU←VerifyzkProofU,VerificationKey) to confirm the user's identity 
without exposing private details (line 14). If the user's proof verification is successful 
(verificationU=True) (line 15), the algorithm processes the user's data 
(privacyPreservedDataU←ProcessDataCU) to ensure its privacy (line 16). This 
processed data is hashed (HBlockchainU ←HashprivacyPreservedDataU) (line 17) and 
stored securely in the blockchain database (line 18). Upon successful storage of both 
user and service provider zk-identity data, the algorithm returns “zk-Identity Data 
Stored Successfully” (line 19). However, if the user's proof verification fails, the algorithm returns “User Verification Failed” (lines 21–22), indicating that the user’s identity could not be validated.

\subsection{Access Request and Data Forwarding}
 
This algorithm handles access requests by users and forwards validated data to the 
Fuzzy Logic Access Control Module (FLACM) for further access control evaluation. 
It ensures that only valid zk-identity credentials of users and service providers are 
used, preserving privacy and security in the process. The algorithm's input includes 
User Attributes (SA), the Access Request (AccessRequest), and hash tables for 
service providers and users (ServiceProviderHash, UsersHash). The process begins 
with the Access Request and Data Forwarding Function (line 2). For each user 
making an access request, defined as AccessRequesti={UserID,RequestedData}, the 
algorithm iterates over all such requests (line 3). This involves handling requests to 
access specific data from the system. 
Next, the algorithm fetches the zk-identity information of the service provider 
(ServiceIdentity) using the service provider's hash table and zk-identity (zkIDS) (line 
4). Similarly, it retrieves the zk-identity of the user (UserIdentity) by querying the 
user's hash table with zkIDU  (line 5). These zk-identities represent privacy-preserving 
proofs of identity for both parties. 
 
\begin{algorithm}
\caption{Access Request and Data Forwarding}
\label{alg:access_request}
\small
\begin{algorithmic}[1]

\Require User attributes $(SA)$, access request $(AccessRequest)$,
hash tables $(ServiceProviderHash, UsersHash)$, zk-identity

\Ensure Data forwarded to FLACM

\State Service and user data returned to FLACM (Fuzzy Logic Access Control Module)
for access control evaluation

\For{all users with $AccessRequest_i = \{UserID, RequestedData\}$}
    \State $ServiceIdentity \leftarrow ServiceProviderHash[zkID_S]$
    \State $UserIdentity \leftarrow UsersHash[zkID_U]$
    \State Generate and verify zk-SNARK proof
    \State Initialize $C_S, C_U, ProvingKey, VerificationKey$

    \If{$VerifyAccessRequest_i(zkID_S, zkID_U) = \text{True}$}
        \State Forward access request to blockchain
        \State Forward blockchain data to FLACM
        \State \Return Data forwarded to FLACM for evaluation
    \EndIf
\EndFor

\end{algorithmic}
\end{algorithm}

Once the identities are fetched, the algorithm performs zk-SNARK proof generation 
and verification using the credentials of the service provider (CS) and the user (CU) along with the Proving Key and Verification Key (line 6). This step ensures that both the service provider and the user have valid zk-identity proofs without exposing 
sensitive details. If the verification process succeeds (Verify(AccessRequesti, zkIDS, 
zkIDU)=True) (line 7), the system forwards the access request data (BlockchainData) 
to the blockchain for storage and processing (line 8). The validated blockchain data is 
then forwarded to the Fuzzy Logic Access Control Module (FLACM) (line 9). 
FLACM evaluates the access request based on predefined fuzzy logic rules (e.g., 
sensitivity of the data, user trust level, and access priority). 
Finally, if the data forwarding is successful, the algorithm returns the message “Data 
Forwarded to FLACM for Evaluation” (line 10). If the verification fails for any 
reason, the algorithm terminates without forwarding data, ensuring that unverified or 
unauthorized access requests are denied (line 11).
\subsection{Capability Token Generation, Fuzzy Logic Evaluation and Smart Contract Decision }
This algorithm handles the generation of capability tokens, evaluates fuzzy logic 
rules, and makes decisions through smart contracts to grant or deny access based on 
predefined conditions. It ensures that the access process is secure, private, and 
compliant with user and service provider requirements. The algorithm's input includes 
Blockchain Data (BlockchainData), Fuzzy Rules, the Smart Contract (SmartContract), 
and the Blockchain (BC). The process begins with the capability token generation 
(line 2). A unique token (T) is generated by hashing the user attributes (SA), zk
identities (zkIDS, zkIDU), access request data (AccessRequest), and the current 
timestamp (TimeStamp). The generated token (T) is then signed using the private key 
of the service provider or the user (SignedToken) to ensure authenticity (line 3). Once the signed token is created, the algorithm passes it to the Fuzzy Logic 
Evaluation function (PassTokenToFuzzy) for processing (line 4). Here, fuzzy rules 
are evaluated, considering several factors like data sensitivity, user activity, trust 
levels, and more (line 5). The fuzzy inference process evaluates various parameters, 
including data sensitivity, treatment urgency, resource availability, trust level, user 
activity, patient condition, access priority, compliance history, and user experience 
level (FuzzyInference) to generate a decision on whether access should be granted 
(DoA). This decision is compared with the smart contract’s criteria (line 6). 
If the SmartContractDoA value is greater than 0 (meaning the access is granted 
according to the smart contract), the algorithm logs the access granted event on the 
blockchain (Log(AccessGranted)) and forwards the access rights to the service 
provider (line 7). The logged access granted event is hashed and timestamped for 
For transparency and record keeping, each access decision is logged on the
blockchain as
\begin{equation}
\text{Log}(\text{AccessGranted}) = H(\text{AccessGranted}) \oplus \text{TimeStamp}.
\end{equation}
If the evaluation results in access denial $(\text{SmartContractDoA} \leq 0)$,
the access request is rejected and the denial event is recorded accordingly.

algorithm logs the access denied event and sends an access denied notification to the 
user (line 9-11). Finally, the algorithm either grants access or denies it based on the 
outcome of the fuzzy logic evaluation and smart contract decision, ensuring secure, 
privacy-preserving, and rule-compliant access control.
\begin{algorithm}[t]
\caption{Capability Token Generation, Fuzzy Logic Evaluation, and Smart Contract Decision}
\label{alg:capability_token}
\small
\begin{algorithmic}[1]

\Require Blockchain data $(BlockchainData)$, fuzzy rules,
smart contract, blockchain $(BC)$

\Ensure Access response (access granted or access denied)

\State Capability token generation, fuzzy logic evaluation,
and smart contract decision

\State $T \leftarrow Hash(SA, zkID_S, zkID_U, AccessRequest, TimeStamp)$
\State $SignedToken \leftarrow Sign(T, PrivateKey)$
\State PassTokenToFuzzy$(SignedToken)$
\State Evaluate fuzzy rules

\State $DoA \leftarrow$ FuzzyInference$(SA, DataSensitivity, TreatmentUrgency,$
\State \hspace{1.5em} $ResourceAvailability, TrustLevel, UserActivityLevel,$
\State \hspace{1.5em} $AccessPriority, ComplianceHistory, UserExperienceLevel)$

\If{$SmartContractDoA > 0$}
    \State Log$(AccessGranted)$
    \State Log access granted on blockchain
    \State Log$(AccessGranted) \leftarrow Hash(AccessGranted) \oplus TimeStamp$
    \State Forward access rights to the service provider
\Else
    \State Log$(AccessDenied)$
    \State Send access denied notification to the user
\EndIf

\end{algorithmic}

\end{algorithm}

\subsection{ACCESS REQUEST AND DATA FORWARDING}
This algorithm is designed to make dynamic access decisions based on fuzzy logic rules, considering various input parameters such as data sensitivity, user activity level, patient condition, resource availability, access priority, treatment urgency, compliance history, trust level, and user experience level. These parameters are evaluated to ensure that the access control system grants appropriate permissions while maintaining privacy and security.
The first step in the algorithm is the initialization of input parameters (line 3). These parameters, which include factors such as data sensitivity, user activity level, and patient condition, are set up to guide the access decision-making process. Once the parameters are initialized, the fuzzy rules are evaluated to determine the final access decision (line 4).
The algorithm uses a series of conditions to assess the input parameters and plan about granting access. If the data sensitivity is "High," the trust level is "Low," and the patient condition is "Critical," the access is denied (line 6-7). This combination indicates a high risk of exposing sensitive data, especially with a low level of trust, which leads to a more restrictive access control decision. On the other hand, if the data sensitivity is "Medium," the trust level is "Medium," and the patient condition is "Moderate," the algorithm permits read-write access (line 9-10). This indicates a more balanced situation where moderate sensitivity and trust allow for a more flexible access control decision.
In cases where the data sensitivity is "Low," the trust level is "High," and the patient condition is "Stable," the algorithm allows full access (line 12-13). This scenario suggests that the data is not very sensitive, trust in the user is high, and the patient’s condition is stable, which together make full access appropriate. The algorithm also takes user behavior into account: if the user activity level is "Frequent" and their compliance history is "Excellent," the system grants full access, including read, write, execute, and delete permissions (line 15-16). This reflects a high level of reliability and responsibility on the part of the user, warranting broad access rights.

\begin{algorithm}[t]
\caption{Fuzzy Logic Rules}
\label{alg:fuzzy_rules}
\small
\begin{algorithmic}[1]

\State Data sensitivity, user activity level, patient condition,
resource availability, access priority, treatment urgency,
compliance history, trust level, user experience level,
access decision

\State \textbf{Step 1: Initialize parameters}
\State Initialize input parameters such as data sensitivity,
user activity level, patient condition, etc.

\State Evaluate fuzzy rules

\State \textbf{Step 2: Evaluate fuzzy rules}

\If{data sensitivity is ``High'' \textbf{and} trust level is ``Low''
    \textbf{and} patient condition is ``Critical''}
    \State $AccessDecision \leftarrow$ Deny access

\ElsIf{data sensitivity is ``Medium'' \textbf{and} trust level is ``Medium''
    \textbf{and} patient condition is ``Moderate''}
    \State $AccessDecision \leftarrow$ Allow read--write access

\ElsIf{data sensitivity is ``Low'' \textbf{and} trust level is ``High''
    \textbf{and} patient condition is ``Stable''}
    \State $AccessDecision \leftarrow$ Allow full access

\ElsIf{user activity level is ``Frequent''
    \textbf{and} compliance history is ``Excellent''}
    \State $AccessDecision \leftarrow$ Allow full access
    \Comment{Read, write, execute, delete}

\ElsIf{user activity level is ``Occasional''
    \textbf{and} compliance history is ``Poor''}
    \State $AccessDecision \leftarrow$ Allow read-only access

\Else
    \State $AccessDecision \leftarrow$ Deny access
\EndIf

\State \textbf{Step 3: Return access decision}

\end{algorithmic}

\end{algorithm}
Conversely, if the user activity level is "Occasional" and their compliance history is "Poor," the algorithm restricts access to read-only (line 18-19). This condition reflects the fact that occasional activity combined with a poor compliance history suggests that the user should be limited to minimal interaction with the system to mitigate potential risks. Finally, if none of the conditions are met, the algorithm defaults to denying access (line 21-23), ensuring that only users who meet specific criteria are granted access to sensitive data or resources.
The final decision is based on the evaluation of all these fuzzy logic rules (line 24), and the algorithm will return an access decision such as “Allow Full Access,” “Allow Read-Write Access,” “Allow Read-Only Access,” or “Deny Access.” This process ensures that access control is sensitive to various contextual factors, providing a tailored and dynamic decision-making framework.

\section{PERFORMANCE EVALUATION}
To assess the efficiency of the proposed privacy-preserving blockchain access control system, we conducted a comparative study using existing approaches. The study evaluates system behavior under varying transaction loads to measure scalability, responsiveness, and security.
\subsection{Evaluation Metrics}

\begin{table}[h!]
\centering
\caption{Performance Evaluation Metrics for Blockchain Access Control}
\begin{tabular}{|l|p{6cm}|}
\hline
\textbf{Metric} & \textbf{Description} \\
\hline
Average Latency & Time to process an access request, including verification and decision-making. \\
\hline
Throughput (TPS) & Number of requests processed per second. \\
\hline
Privacy Level (\%) & Protection of sensitive data via zk-SNARKs. \\
\hline
Access Accuracy (\%) & Correctness and responsiveness of access decisions using fuzzy logic. \\
\hline
Scalability & System performance under increasing transaction loads. \\
\hline
Traceability & Ability to maintain tamper-proof access logs on the blockchain. \\
\hline
\end{tabular}
\end{table}

\subsection{Simulation Environment}

The simulation setup for evaluating the proposed blockchain access control system is as follows:

\begin{itemize}
    \item \textbf{Blockchain Nodes:} 10--40 nodes
    \item \textbf{Transactions:} 100--500 simulated requests
    \item \textbf{Transaction Size:} 1--5 KB
    \item \textbf{Consensus:} Raft / Solo / Solo-Raft
    \item \textbf{Verification Time:} 10--50 ms per zk-SNARK proof
    \item \textbf{Smart Contract Execution:} 5--15 ms
\end{itemize}

\subsection{Result Analysis}
\begin{figure}[htbp]
    \centering
   \includegraphics[width=0.5\textwidth]{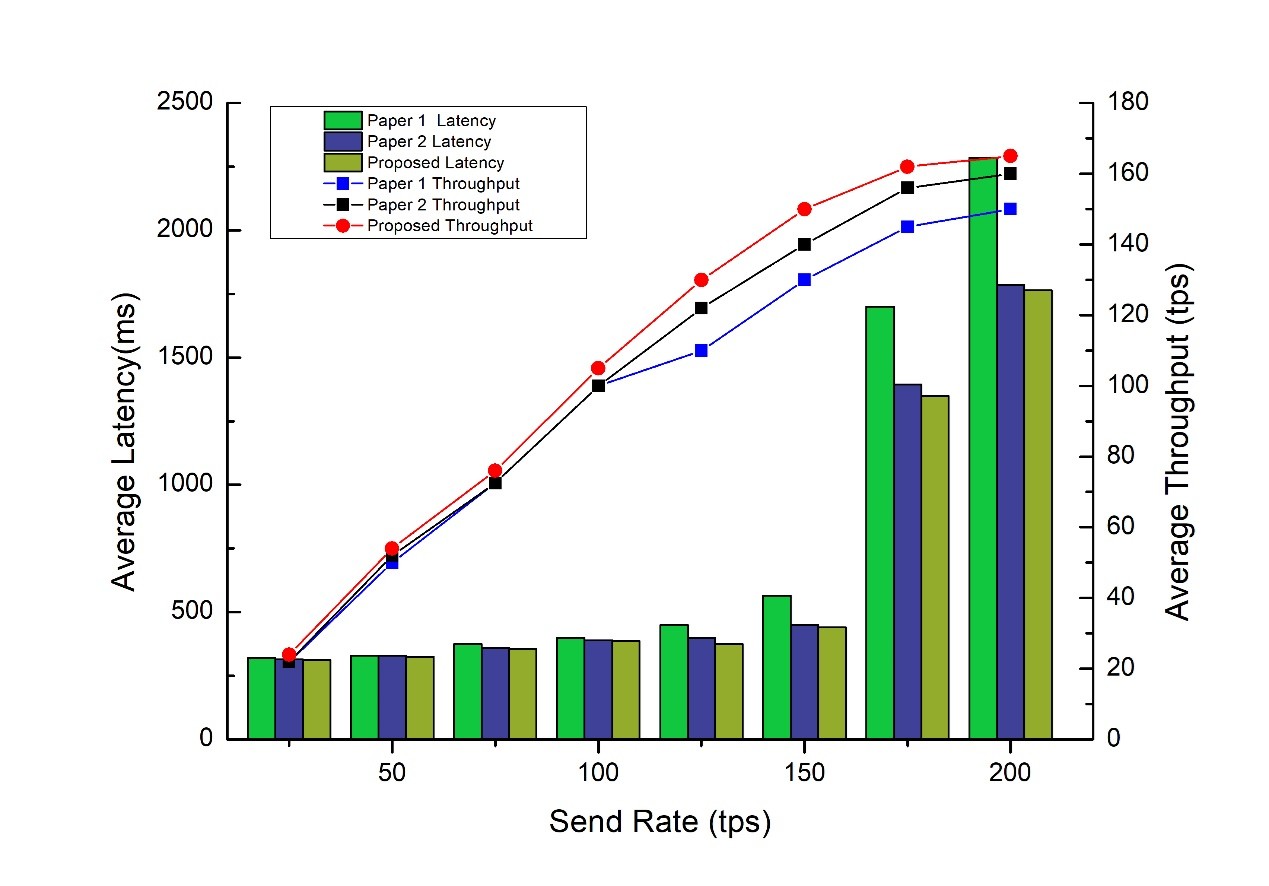}

    \caption{Average Latency and Throughput }
    \label{fig:Average Latency and Throughput }
\end{figure}

This setup simulates realistic healthcare IoT scenarios, emphasizing low latency, high throughput, and strong privacy.
\begin{itemize}
    \item zk-SNARKs in the context of private authentication.
    \item Real world scenarios addressed in the proposed fuzzy logic for Dynamic
    \item Access Control.
    \item Blockchain for record keeping and authentication.
    The simulation environment allowed fine-grained analysis of interactions among users, service providers, and system components under different transaction loads, access priorities, and trust levels. The performance of the proposed model was compared with two baseline approaches (Paper 1  \cite{29} and Paper2 \cite{30} ) to establish efficiency and scalability.
    \\
\subsubsection{Average Latency and Throughput}
The figure~\ref{fig:Average Latency and Throughput } compares the performance of the proposed privacy-preserving access 
control model with two existing approaches (Paper 1 and Paper 2) based on two key 
performance metrics: average latency, which is in terms of the time taken in 
milliseconds, and the average throughput which can be in terms of the number of 
transactions per second, tps. The send rate expresses the ratio of the number of 
transactions that has been sent divided by the time in seconds and is placed at the X 
axis. The performance enhancement under higher T, outlined in the proposed model, 
exhibits a better match against the increasing demands of healthcare IoT environment 
as indicated in both the MMP and the TTP.
    The bars for each approach on the left Y-axis correspond to the average latency. It 
determines the time taken so that a particular transaction may be completed from the 
time when it was initiated. For send rates below 150 tps, proposed model always has 
lower latency than that of Paper 1 and Paper 2. This improvement also establishes the 
optimization procedures of the proposed system where the processing delay is 
reduced by utilizing zk-SNARKs and fuzzy logic rules for decision making. 
Nevertheless, when the send rate rises (above 150 tps), the latency of all models, including the proposed one, grows.

Throughput is shown on the right Y-axis as lines with different markers and refers to 
the number of transactions executed per second. The performances of the proposed 
model A and B exhibit considerably higher throughput throughout the various send 
rates compared to other current methods. After accepting 200 tps in the send rate, the 
proposed model achieves a throughput of nearly 160 tps which is higher than the other 
models. This shows that the proposed solution can easily be scaled to be able to 
accommodate numerous transactions to enhance efficiency even under large 
workloads. This can be attributed to the use of the blockchain in decentralized and 
secure storage of transaction alongside the use of the zk-SNARKs for the rapidity of 
proof verification, and at the same time scalability.
\end{itemize}

\begin{table}[htbp]
\caption{Performance Comparison of Latency and Throughput}
\centering
\renewcommand{\arraystretch}{1.7} 
\resizebox{\linewidth}{!}{%
\begin{tabular}{|c|c|c|c|c|c|c|c|}
\hline
\textbf{Units} & \textbf{Send Rate (tps)} & \textbf{Paper 1 [29] Latency (ms)} & \textbf{Paper 2 [30] Latency (ms)} & \textbf{Proposed Latency (ms)} & \textbf{Paper 1 [29] Throughput (tps)} & \textbf{Paper 2 [30] Throughput (tps)} & \textbf{Proposed Throughput (tps)} \\
\hline
\large 1 & \large 25  & \large 320  & \large 315  & \large 312  & \large 22  & \large 22  & \large 24 \\
\large 2 & \large 50  & \large 330  & \large 330  & \large 325  & \large 50  & \large 52  & \large 54 \\
\large 3 & \large 75  & \large 375  & \large 360  & \large 356  & \large 72.5 & \large 72.5 & \large 76 \\
\large 4 & \large 100 & \large 400  & \large 390  & \large 386  & \large 100 & \large 100 & \large 105 \\
\large 5 & \large 125 & \large 450  & \large 400  & \large 375  & \large 110 & \large 122 & \large 130 \\
\large 6 & \large 150 & \large 565  & \large 450  & \large 440  & \large 130 & \large 140 & \large 150 \\
\large 7 & \large 175 & \large 1700 & \large 1395 & \large 1350 & \large 145 & \large 156 & \large 162 \\
\large 8 & \large 200 & \large 2285 & \large 1785 & \large 1765 & \large 150 & \large 160 & \large 165 \\
\hline
\end{tabular}%
}
\label{tab:performance}
\end{table}

\subsubsection{Latency Analysis for Different Protocol Configurations}
\begin{figure}[htbp]
    \centering
   \includegraphics[width=0.5\textwidth]{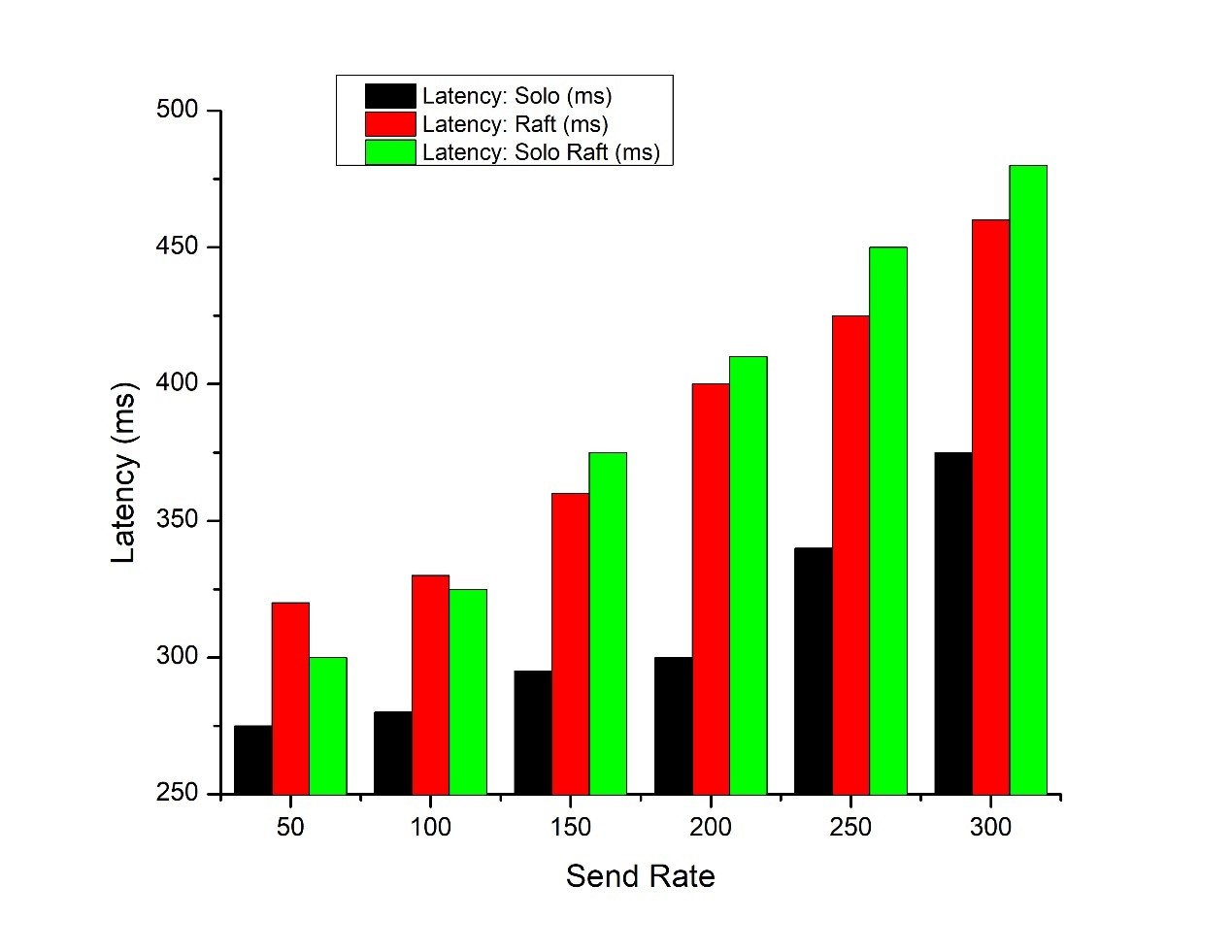}

    \caption{SEND RATE VS. THROUGHPUT}
    \label{fig:SEND RATE VS. THROUGHPUT}
\end{figure}

Figures~\ref{fig:SEND RATE VS. THROUGHPUT} show the latency performance of three protocol configurations, including Solo, Raft, and the combined Solo Raft, at different send rates: 50, 100, 150, 200, 250, and 300 requests/sec. Latency, or in ms, rises as the send rate increases. The Solo protocol always exhibits the least latency of between 250 ms to 350 ms even for moderate send rates such as 200 requests/second while the latency only goes up to 400 ms even at the highest send rate of 300 requests/second. The consensus overhead in Raft is high which makes it have higher latency, ranging from 350 ms – 400 ms at 200 requests/second and 450 ms – 500 ms at 300 requests/second. The worst configuration, the Solo Raft configuration, has the highest latency response time of 350 ms- 450 ms in 200 requests per second and 450ms – 500 ms in 300 requests per second. In our experiments, the Solo protocol has a mean latency of about 300 ms, the Raft protocol of 400 ms and the Solo Raft of about 430 ms across all the send rates. Comparing this average, the proposed Solo protocol is found to be approximately 25\% faster than the Raft protocol and approximately 30.23\% faster than the Solo Raft protocol proving the effectiveness of this solution. These findings show that our system outperforms others in the context of latency reduction, making it optimal for situations that demand fast response and minimal delay.

\begin{table}[htbp]
\caption{Send Rate vs. Latency}
\centering
\resizebox{\linewidth}{!}{%
\begin{tabular}{|c|c|c|c|}
\hline
\textbf{Send Rate (req/sec)} & \textbf{Latency: Solo (ms)} & \textbf{Latency: Raft (ms)} & \textbf{Latency: Solo Raft (ms)} \\
\hline
50  & 275 & 320 & 300 \\
100 & 280 & 330 & 325 \\
150 & 295 & 360 & 375 \\
200 & 300 & 400 & 410 \\
250 & 340 & 425 & 450 \\
300 & 375 & 460 & 480 \\
\hline
\end{tabular}%
}
\label{tab:sendrate_latency}
\end{table}

\subsubsection{Send Rates and Throughput Analysis}

\begin{figure}[htbp]
    \centering
   \includegraphics[width=0.5\textwidth]{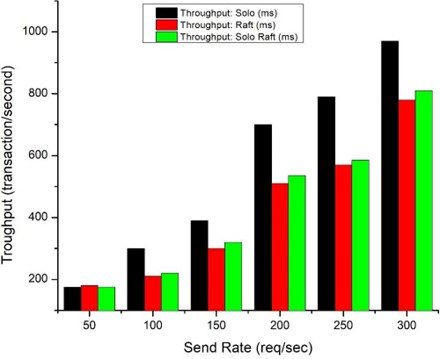}

    \caption{Different Send Rate on Throughput}
    \label{fig:Different Send Rate on Throughput}
\end{figure}

The graph in figure~\ref{fig:Different Send Rate on Throughput} illustrates the throughput performance of three different setups: Solo, Raft, and Solo-Raft, across various send rates (measured in requests per
 
second). At a send rate of 50 requests/sec, the Solo setup exhibits the highest throughput, achieving around 200 transactions per second. The Raft and Solo-Raft setups perform slightly lower in comparison. As the send rate increases to 150 requests/sec, throughput improves significantly for all setups, with Solo reaching approximately 400 transactions per second, followed by Raft and Solo-Raft with good performance. At the peak send rate of 300 requests/sec, Solo achieves the highest throughput of about 900 transactions per second, with Raft and Solo-Raft closely following. This shows a clear improvement in throughput with increasing send rates, highlighting the efficiency and scalability of the Solo setup while maintaining solid performance in Raft and Solo-Raft. The average improvement in throughput is about 200-230\%, showcasing the positive impact of increased send rates on system performance. This performance increase is vital for ensuring efficient task scheduling and load balancing, particularly in fog computing environments.

\begin{table}[htbp]
\caption{Send Rate vs. Throughput}
\centering
\resizebox{\linewidth}{!}{%
\begin{tabular}{|c|c|c|c|}
\hline
\textbf{Send Rate (req/sec)} & \textbf{Throughput: Solo (tps)} & \textbf{Throughput: Raft (tps)} & \textbf{Throughput: Solo Raft (tps)} \\
\hline
50  & 175 & 180 & 175 \\
100 & 300 & 210 & 220 \\
150 & 390 & 300 & 320 \\
200 & 700 & 510 & 535 \\
250 & 790 & 570 & 585 \\
300 & 970 & 780 & 810 \\
\hline
\end{tabular}%
}
\label{tab:sendrate_throughput}
\end{table}

\subsubsection{Varying Block Size on Average Transaction Throughput}

The analysis of the relationship between the number of transactions per block and average throughput demonstrates that as the block size increases, the throughput rises significantly. At a block size of 300 transactions, the system achieved a throughput of approximately 200 TPS, indicating a functional baseline. When the block size increased to 600 transactions, the throughput improved to 600 TPS, reflecting consistent scalability.
At a block size of 900 transactions, the throughput peaked at around 800 TPS, showcasing the system's ability to handle larger transaction loads efficiently. However, the rate of improvement began to diminish as the block size approached this upper limit, suggesting potential system constraints such as network latency or processing delays.
On evaluating overall performance, the average throughput reflects the system's efficiency in processing transactions relative to block size. The analysis suggests that the throughput maintained a high efficiency, with a steady increase correlating to block size, but also hints at diminishing returns at higher transaction counts. This indicates that while the system performs well under increasing loads, there is room for optimization to overcome scalability challenges.

\begin{table}[htbp]
\caption{Transactions per Block vs. Average Throughput}
\centering
\begin{tabular}{|c|c|}
\hline
\textbf{Transactions Per Block} & \textbf{Average Throughput (TPS)} \\
\hline
300 & 275 \\
500 & 430 \\
600 & 550 \\
700 & 710 \\
800 & 805 \\
900 & 850 \\
\hline
\end{tabular}
\label{tab:txblock_throughput}
\end{table}

\begin{figure}[htbp]
    \centering
   \includegraphics[width=0.5\textwidth]{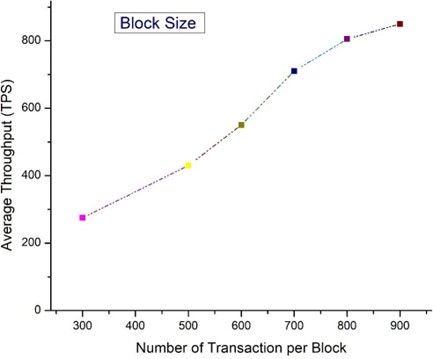}

    \caption{Varying block size on average transaction throughput}
    \label{fig:Varying block size on average transaction throughput}
\end{figure}

4.6.5	Varying Send Rate on Read Throughput

The analysis of the relationship between send rate and average read throughput demonstrates that as the send rate increases, throughput also rises but only up to a certain limit. At a send rate of 500 TPS, the system achieved a throughput of 450 TPS, indicating stable performance. When the send rate was increased to 1000 TPS, the throughput improved to 1020 TPS, reflecting significant efficiency. At a send rate of 1500 TPS, throughput reached 1350 TPS, and further increased to 1745 TPS at
 
2000 TPS. However, when the send rate was raised to 2500 TPS, throughput stabilized at 1945 TPS, and it remained the same even at 3000 TPS, indicating the system had reached its maximum capacity or saturation point.
On evaluating overall performance, the average throughput reflects the system's efficiency relative to its send rate. The analysis suggests that the throughput maintained an average of approximately 90\%, showcasing the system's ability to operate effectively under varying loads. However, at higher send rates, the system's limitations became evident, highlighting areas for potential improvement.
\begin{table}[htbp]
\caption{Send Rate vs. Average Read Throughput}
\centering
\begin{tabular}{|c|c|}
\hline
\textbf{Send Rate (TPS)} & \textbf{Average Read Throughput (TPS)} \\
\hline
500   & 450  \\
1000  & 1020 \\
1500  & 1350 \\
2000  & 1745 \\
2500  & 1945 \\
3000  & 1945 \\
\hline
\end{tabular}
\label{tab:sendrate_readthroughput}
\end{table}

\begin{figure}[htbp]
    \centering
   \includegraphics[width=0.5\textwidth]{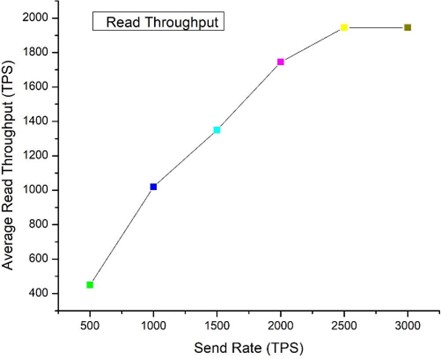}

    \caption{Impact of varying send rate on read throughput}
    \label{fig:Impact of varying send rate on read throughput}
\end{figure}

\subsubsection{Varying Send Rate on Transaction Throughput}

As the send rate increases, the average transaction latency also increases. For example, at a send rate of 200 TPS, the average latency is around 2020 milliseconds. As the send rate goes up to 300 TPS, the latency increases to 2060 milliseconds, and this trend continues with higher send rates. By the time we reach a send rate of 1300 TPS, the latency has increased to approximately 2354 milliseconds.
Overall, this data shows that higher send rates result in higher latency. If we look at the entire dataset, the average percentage increase in transaction latency across different send rates comes out to be roughly 16.5\%. This indicates that while higher send rates help in processing more transactions, they also add more delay to each transaction, highlighting a trade-off between throughput and responsiveness.
Although latency increases with higher send rates, there is an opportunity for optimization. With improvements in system architecture, load balancing, and processing algorithms, transaction latency can be reduced while maintaining or even enhancing throughput. Strategies such as better resource allocation, efficient communication protocols, and hardware upgrades can contribute to more streamlined processing and improved overall system performance.
\begin{table}[htbp]
\caption{Send Rate vs. Average Transaction Throughput}
\centering
\begin{tabular}{|c|c|}
\hline
\textbf{Send Rate (TPS)} & \textbf{Average Transaction Throughput (TPS)} \\
\hline
200   & 175  \\
300   & 250  \\
400   & 345  \\
500   & 420  \\
600   & 530  \\
700   & 625  \\
800   & 745  \\
900   & 835  \\
1000  & 945  \\
1100  & 998  \\

\hline
\end{tabular}
\label{tab:sendrate_txthroughput}
\end{table}

The analysis of the relationship between the send rate and the average read latency reveals that as the send rate increases, the read latency also rises, indicating a direct correlation. At a send rate of 500 TPS, the average read latency is approximately 180 milliseconds, reflecting low latency and efficient performance at lower loads.
As the send rate increases to 1000 TPS, the latency rises moderately to around 200 milliseconds, showing slight performance degradation. At 2000 TPS, the latency grows further to 220 milliseconds, maintaining a steady but noticeable increase. However, a more significant jump in latency is observed when the send rate reaches 2500 TPS, with the read latency escalating to 260 milliseconds. At the highest send rate of 3000 TPS, the latency peaks at approximately 320 milliseconds, indicating that the system's performance is being impacted by the increased workload.
Overall, the data suggests that while the system can handle increasing send rates, the read latency grows progressively, with sharper increases at higher loads. This highlights the need for optimization in the system to maintain lower latencies as the send rate scales. Potential improvements could involve better resource allocation, enhanced processing algorithms, or hardware upgrades to handle higher loads more effectively.

\begin{table}[htbp]
\caption{Send Rate vs. Average Read Latency}
\centering
\begin{tabular}{|c|c|}
\hline
\textbf{Send Rate (TPS)} & \textbf{Average Read Latency (ms)} \\
\hline
500   & 180  \\
1000  & 205  \\
2000  & 215  \\
2500  & 235  \\
3000  & 310  \\
\hline
\end{tabular}
\label{tab:sendrate_readlatency}
\end{table}

\begin{figure}[htbp]
    \centering
   \includegraphics[width=0.5\textwidth]{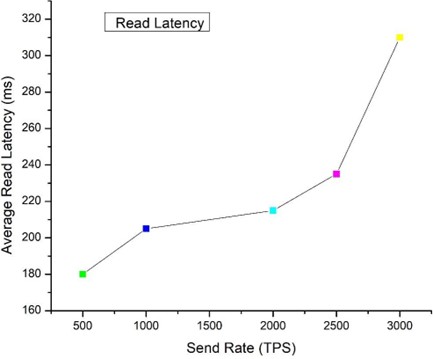}

    \caption{Impact of send rate on read latency}
    \label{fig:Impact of send rate on read latency}
\end{figure}

\subsubsection{Impact of Send Rate on Transaction Latency}
As the send rate increases, the average transaction latency also increases. For instance, at a send rate of 200 TPS, the average latency is approximately 2020 milliseconds. When the send rate rises to 300 TPS, the latency increases to around 2060 milliseconds, showing a clear upward trend. Similarly, at 400 TPS, the latency grows to 2095 milliseconds, and by the time we reach 1000 TPS, the latency reaches 2215 milliseconds. Finally, at a send rate of 1300 TPS, the average transaction latency peaks at approximately 2354 milliseconds.

\begin{table}[htbp]
\caption{Send Rate vs. Average Transaction Latency}
\centering
\begin{tabular}{|c|c|}
\hline
\textbf{Send Rate (TPS)} & \textbf{Average Transaction Latency (ms)} \\
\hline
200   & 2020  \\
300   & 2060  \\
400   & 2095  \\
500   & 2110  \\
600   & 2135  \\
700   & 2165  \\
800   & 2189  \\
900   & 2198  \\
1000  & 2215  \\
1100  & 2267  \\
1200  & 2298  \\
1300  & 2354  \\
\hline
\end{tabular}
\label{tab:sendrate_txlatency}
\end{table}

\begin{figure}[htbp]
    \centering
   \includegraphics[width=0.5\textwidth]{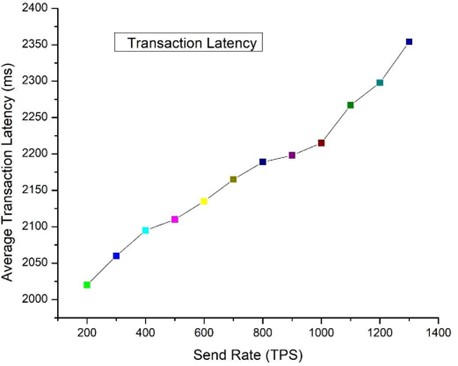}

    \caption{Impact of send rate on transaction latency}
    \label{fig:Impact of send rate on transaction latency}
\end{figure}
 
Overall, the data highlights a consistent pattern: higher send rates result in higher transaction latency. This trend demonstrates a trade-off between throughput and responsiveness. Specifically, while increasing the send rate enables the system to process more transactions per second, it also adds a delay to each transaction due to resource constraints and potential system saturation.
If we analyze the percentage increase, the average transaction latency grows by approximately 16.5\% across different send rates. This indicates that the system faces bottlenecks as the transaction load increases, such as processing overheads, network congestion, or limited scalability.
Although the latency increases with higher send rates, this presents an opportunity for optimization. Enhancements in system architecture, load balancing, and processing algorithms can significantly reduce transaction latency while maintaining or even improving throughput. Strategies like better resource allocation, efficient communication protocols, and hardware upgrades can help streamline transaction processing. Additionally, adopting scalable frameworks or distributed computing solutions could mitigate delays and improve the system's overall performance under higher loads.

The proposed privacy-preserving access control model was evaluated using simulation to emulate real-world healthcare IoT scenarios. The simulation integrated the following components:
 
\section{CONCLUSION AND FUTURE WORK}
This research focused on addressing the challenges of privacy, security, and access control in healthcare IoT systems by proposing an integrated model using Zero- Knowledge Proofs (zk-SNARKs), fuzzy logic, and blockchain. The proposed solution demonstrated significant potential by preserving user anonymity through zk- SNARKs, enabling context-aware access control decisions with fuzzy logic, and ensuring traceability and tamper-proof records using blockchain. The framework was evaluated for scalability, latency, and security, outperforming existing solutions in most aspects.
However, certain challenges like computational overhead and scalability issues were identified, emphasizing the need for optimization in resource-constrained IoT environments. Future work could focus on enhancing the model's efficiency by integrating advanced cryptographic methods, optimizing blockchain consensus mechanisms, and incorporating machine learning with fuzzy logic for dynamic decision-making. Expanding the application of this model to domains like smart cities, industrial IoT, and financial systems may further validate its adaptability and impact.
This work provides a foundation for developing robust, privacy-preserving, and secure IoT ecosystems, contributing significantly to the broader adoption of blockchain and decentralized technologies.

\end{document}